\documentstyle[12pt]{article}
\topmargin - 1.0 true cm
\newcommand{\beq}{\begin{equation}}
\newcommand{\eeq}{\end{equation}}
\def\sss{\scriptscriptstyle}

\def\bd{B_d^0}
\def\bdbar{{\overline{B_d^0}}}
\def\bs{B_s^0}
\def\bsbar{{\overline{B_s^0}}}
\def\barp{{\raise.35ex\hbox{${\sss (}$}}---{\raise.35ex\hbox{${\sss )}$}}}
\def\bdbarp{\hbox{$B_d$\kern-1.4em\raise1.4ex\hbox{\barp}}}
\def\bsbarp{\hbox{$B_s$\kern-1.4em\raise1.4ex\hbox{\barp}}}
\def\ks{K_{\sss S}}
\def\roughly#1{\mathrel{\raise.3ex\hbox{$#1$\kern-.75em\lower1ex\hbox{$\sim$}}}}
\def\lsim{\roughly<}

\def\npb#1#2#3{{\it Nucl.\ Phys.} {\bf B#1} (19#2) #3}
\def\plb#1#2#3{{\it Phys.\ Lett.} {\bf #1B} (19#2) #3}
\def\prd#1#2#3{{\it Phys.\ Rev.} {\bf D#1} (19#2) #3}

\def\prl#1#2#3{{\it Phys.\ Rev.\ Lett.} {\bf #1} (19#2) #3}
\def\zpc#1#2#3{{\it Zeit.\ Phys.} {\bf C#1} (19#2) #3}
\def \stone{{\it B Decays}, edited by S. Stone (World Scientific, Singapore,
1994)}
\newread\epsffilein 
\newif\ifepsffileok 
\newif\ifepsfbbfound 
\newif\ifepsfverbose 
\newdimen\epsfxsize 
\newdimen\epsfysize 
\newdimen\epsftsize 
\newdimen\epsfrsize 
\newdimen\epsftmp 
\newdimen\pspoints 
\def\epsfbox#1{\global\def\epsfllx{72}\global\def\epsflly{72}%
 \global\def\epsfurx{540}\global\def\epsfury{720}%
 \def\lbracket{[}\def\testit{#1}\ifx\testit\lbracket
 \let\next=\epsfgetlitbb\else\let\next=\epsfnormal\fi\next{#1}}%
\def\epsfgetlitbb#1#2 #3 #4 #5]#6{\epsfgrab #2 #3 #4 #5 .\\%
 \epsfsetgraph{#6}}%
\def\epsfnormal#1{\epsfgetbb{#1}\epsfsetgraph{#1}}%
\def\epsfgetbb#1{%
\openin\epsffilein=#1
\ifeof\epsffilein\errmessage{I couldn't open #1, will ignore it}\else
 {\epsffileoktrue \chardef\other=12
 \def\do##1{\catcode`##1=\other}\dospecials \catcode`\ =10
 \loop
 \read\epsffilein to \epsffileline
 \ifeof\epsffilein\epsffileokfalse\else
 \expandafter\epsfaux\epsffileline:. \\%
 \fi
 \ifepsffileok\repeat
 \ifepsfbbfound\else
 \ifepsfverbose\message{No bounding box comment in #1; using defaults}\fi\fi
 }\closein\epsffilein\fi}%
\def\epsfclipstring{}
\def\epsfsetgraph#1{%
 \epsfrsize=\epsfury\pspoints
 \advance\epsfrsize by-\epsflly\pspoints
 \epsftsize=\epsfurx\pspoints
 \advance\epsftsize by-\epsfllx\pspoints
 \epsfxsize\epsfsize\epsftsize\epsfrsize
 \ifnum\epsfxsize=0 \ifnum\epsfysize=0
 \epsfxsize=\epsftsize \epsfysize=\epsfrsize
 \epsfrsize=0pt
 \else\epsftmp=\epsftsize \divide\epsftmp\epsfrsize
 \epsfxsize=\epsfysize \multiply\epsfxsize\epsftmp
 \multiply\epsftmp\epsfrsize \advance\epsftsize-\epsftmp
 \epsftmp=\epsfysize
 \loop \advance\epsftsize\epsftsize \divide\epsftmp 2
 \ifnum\epsftmp>0
 \ifnum\epsftsize<\epsfrsize\else
 \advance\epsftsize-\epsfrsize \advance\epsfxsize\epsftmp \fi
 \repeat
 \epsfrsize=0pt
 \fi
 \else \ifnum\epsfysize=0
 \epsftmp=\epsfrsize \divide\epsftmp\epsftsize
 \epsfysize=\epsfxsize \multiply\epsfysize\epsftmp
 \multiply\epsftmp\epsftsize \advance\epsfrsize-\epsftmp
 \epsftmp=\epsfxsize
 \loop \advance\epsfrsize\epsfrsize \divide\epsftmp 2
 \ifnum\epsftmp>0
 \ifnum\epsfrsize<\epsftsize\else
 \advance\epsfrsize-\epsftsize \advance\epsfysize\epsftmp \fi
 \repeat
 \epsfrsize=0pt
 \else
 \epsfrsize=\epsfysize
 \fi
 \fi
 \ifepsfverbose\message{#1: width=\the\epsfxsize, height=\the\epsfysize}\fi
 \epsftmp=10\epsfxsize \divide\epsftmp\pspoints
 \vbox to\epsfysize{\vfil\hbox to\epsfxsize{%
 \ifnum\epsfrsize=0\relax
 \includegraphics{#1}%
 \else
 \epsfrsize=10\epsfysize \divide\epsfrsize\pspoints
 \includegraphics{#1}%
 \fi
 \hfil}}%
\global\epsfxsize=0pt\global\epsfysize=0pt}%
 {\catcode`\%=12 \global\let\epsfpercent=
\long\def\epsfaux#1#2:#3\\{\ifx#1\epsfpercent
 \def\testit{#2}\ifx\testit\epsfbblit
 \epsfgrab #3 . . . \\%
 \epsffileokfalse
 \global\epsfbbfoundtrue
 \fi\else\ifx#1\par\else\epsffileokfalse\fi\fi}%
\def\epsfempty{}%
\def\epsfgrab #1 #2 #3 #4 #5\\{%
\global\def\epsfllx{#1}\ifx\epsfllx\epsfempty
 \epsfgrab #2 #3 #4 #5 .\\\else
 \global\def\epsflly{#2}%
 \global\def\epsfurx{#3}\global\def\epsfury{#4}\fi}%
\def\epsfsize#1#2{\epsfxsize}
\begin{document}
\baselineskip=6truemm
\begin{flushright}
UdeM-GPP-TH-97-40 \\
BNL-~~~~~~~~~~~~~~~\\
April, 1997\\
hep-ph/97xxxxx 
\end{flushright}
\bigskip
\begin{center}
{\bf 
Measuring the CP Angle $\beta$ in Hadronic $b\to s$ Penguin Decays}
\medskip\\
David London$^a$
and Amarjit Soni$^b$
\end{center}

\bigskip

\begin{flushleft}
$a$) Laboratoire de physique nucl\'eaire, Universit\'e de Montr\'eal,
C.P. 6128, succ. centre-ville, Montr\'eal, QC, Canada \\
$b$) Theory Group, Brookhaven National Laboratory, Upton, NY\ \ 11973
\end{flushleft}
\bigskip

\begin{quote}
{\bf Abstract}: 
Asymmetric $e^+e^-$ colliders running on the $\Upsilon(4S)$ ($B$ factories)
will much more readily measure CP-violating asymmetries in the decays of
$\bd$ and $B^\pm$ mesons than in the decays of $\bs$ mesons. As such, they
will seemingly not be able to probe new phases in $\bs$-$\bsbar$ mixing,
i.e.\ in $b\to s$ transitions. However, by measuring the CP angle $\beta$
via $b \to s$ hadronic penguin decays such as $\bdbarp \to \eta'\ks$ and
$\bdbarp\to\phi\ks$, and comparing its value to that obtained in
$\bdbarp\to\Psi\ks$, it is possible to detect the presence of new physics
in the $b\to s$ flavour-changing neutral current. Recent CLEO results are
encouraging in this regard. They suggest that the branching ratio of the
$b\to s$ penguin decay $B_d^0 \to \eta'\ks$ is anomalously large, about $4
\times 10^{-5}$, which will make it much easier to search for new physics
in $b\to s$ transitions.
\end{quote}
\newpage

\begin{figure}
\vskip -1.0truein
\centerline{\epsfxsize 3.5 truein \epsfbox {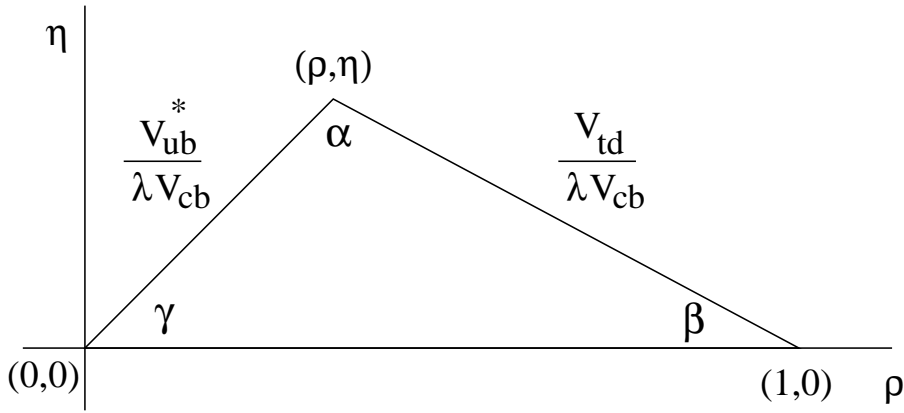}}
\vskip -1.2truein
\caption{The unitarity triangle. The angles $\alpha$, $\beta$ and $\gamma$
can be measured via CP violation in the $B$ system.}
\label{triangle}
\end{figure}

To date, the only experimental evidence for CP violation in the weak
interactions comes from the kaon system, where CP violation in
$K^0$-${\overline{K^0}}$ mixing has been observed. According to the
standard model (SM), this CP violation is due to a complex phase in the
Cabibbo-Kobayashi-Maskawa (CKM) matrix. Within the Wolfenstein
parametrization of the CKM matrix \cite{Wolfenstein}, the only elements
which have non-negligible phases are $V_{td}$ and $V_{ub}$:
\beq
V_{\sss CKM} = \left(
\matrix{ 1-{1\over 2}\lambda^2 & \lambda & A \lambda^3 (\rho - i \eta) \cr
- \lambda & 1-{1\over 2}\lambda^2 & A \lambda^2 \cr
A \lambda^3 (1 - \rho - i\eta) & - A \lambda^2 & 1 \cr} \right),
\eeq
where $\lambda = 0.22$ is the Cabibbo angle. These two complex matrix
elements are conventionally parametrized as $V_{td} \equiv |V_{td}|
\exp(-i\beta)$ and $V_{ub} \equiv |V_{ub}| \exp(-i\gamma)$. The phase
information in the CKM matrix can be elegantly displayed using the
well-known unitarity triangle (Fig.~\ref{triangle}), which is due to the
orthogonality of the first and third columns of the CKM matrix.

In the coming years, this explanation of CP violation will be tested at $B$
factories with the BABAR, BELLE and CLEO detectors. The three angles of the
unitarity triangle, $\alpha$, $\beta$ and $\gamma$ can be extracted through
the measurements of CP-violating asymmetries in $B$ decays to hadronic
final states \cite{CPreview}. For example, the CP asymmetries in
$\bdbarp\to\pi^+\pi^-$ and $\bdbarp\to\Psi\ks$ probe $\alpha$ and $\beta$,
respectively, and the angle $\gamma$ can be extracted from the CP asymmetry
in $B^\pm \to D K^\pm$ \cite{growyler}. In all cases, the CP phases can be
obtained with virtually no hadronic uncertainty. (For the extraction of
$\alpha$, since penguins are unlikely to be negligible, an isospin analysis
(for the $2\pi$ final state \cite{mgdl1}) and/or a Dalitz plot analysis
($3\pi$ final state \cite{hqas}) will probably be necessary.) Of course,
there are many other decays which can be used to obtain the CP angles, but
those mentioned above are the ones which are most often discussed -- they
have become the ``standard'' decay modes.

One alternative way of measuring $\gamma$ is via the CP asymmetry in 
$\bsbarp\to D_s^\pm K^\mp$ \cite{ADK}. However, since in all probability
the $B$ factories are not going to be able to measure CP asymmetries in
$B_s^0$ decays, this method will not be available. This will be an
important point in the following discussion.

If there is new physics, there are basically three ways in which it can
show up in measurements of CP asymmetries \cite{newphysics}:
\begin{itemize}

\item The relation $\alpha+\beta+\gamma=\pi$ is violated.

\item Although $\alpha+\beta+\gamma=\pi$, one finds values for the CP
phases which are outside of the SM predictions.

\item The CP angles measured are consistent with the SM predictions,
and add up to $180^\circ$, but are inconsistent with the measurements of
the {\it sides} of the unitarity triangle.

\end{itemize}
The principal way in which physics beyond the SM affects the CP asymmetries
is via new contributions to $B^0$-${\overline{B^0}}$ mixing. There are, in
fact, many models of new physics which can significantly affect this mixing
\cite{newphysics}. However, because $B$ factories will not measure CP
asymmetries involving $B_s^0$ mesons, the measurements will be sensitive
{\it only} to new physics in $\bd$-$\bdbar$ mixing, i.e.\ in the $b\to d$
flavor-changing neutral current (FCNC). This has the consequence that if,
as is expected, $\alpha$ and $\beta$ are measured using CP asymmetries
involving $B_d^0$ decays, and $\gamma$ is obtained via  $B^\pm \to D
K^\pm$, the $B$ factories will {\it automatically} find that $\alpha +
\beta + \gamma = \pi$ \cite{NirSilv}. This is because any new-physics
effects in $\bd$-$\bdbar$ mixing cancel when $\alpha$ and $\beta$ are
added, and there are no mixing effects in the measurement of $\gamma$. If,
on the other hand, $\gamma$ were measured using $B_s^0$ decays, it would be
possible to find $\alpha + \beta + \gamma \ne \pi$ if there were new phases
in $\bs$-$\bsbar$ mixing, i.e.\ in the $b\to s$ FCNC.

Now, it is conceivable that the new physics significantly affects the $b\to
s$ FCNC, without affecting the $b\to d$ FCNC appreciably. It would be a
shame if this possibility remained untested at the $B$ factories. Is it
really impossible to detect any new phases appearing in $\bs$-$\bsbar$
mixing? 

Fortunately, the answer to this question is no. It is in fact possible to
detect new phases in the $b\to s$ FCNC, but in a somewhat different way.
One of the alternative possibilities for measuring the angle $\beta$
involves $b \to s$ penguin decays \cite{LonPeccei}. Such decays, being
dominated by virtual $t$-quarks, involve the combination of CKM matrix
elements $V_{tb}^* V_{ts}$, which is real in the Wolfenstein
parametrization. Thus, in the context of the SM, if one measures the CP
asymmetry in $\bdbarp\to f$, where $f$ is a CP eigenstate and the decay is
dominated by a $b \to s$ hadronic penguin, one is probing the phase of
$\bd$-$\bdbar$ mixing, which is simply $\beta$. 

The key point here is that this is the same CKM phase as that probed in
$\bdbarp\to\Psi\ks$. Therefore, if one finds a discrepancy between the
values of $\beta$ as measured in $\bdbarp\to\Psi\ks$ and in hadronic $b\to
s$ penguins, this clearly points to new physics, with new phases, in the
$b\to s$ FCNC. In particular, it indicates that there are new amplitudes
contributing to these hadronic $b\to s$ penguin decays \cite{GrossWor}.
This same new physics will, in general, also lead to new phases in
$\bs$-$\bsbar$ mixing, and hence to disagreements with the SM predictions
for CP asymmetries in $\bs$ decays. Thus, by measuring $\beta$ in hadronic
$b\to s$ penguins, one is performing essentially the same tests as if one
measured CP asymmetries in $\bs$ decays. (Note that this holds even if
there is new physics in $\bd$-$\bdbar$ mixing.)

There are several models of new physics which can lead to new phases in the
$b\to s$ FCNC \cite{newphysics,GrossWor}: four generations, non-minimal
supersymmetric models such as effective supersymmetry \cite{effSUSY}, and
models with enhanced chromomagnetic dipole operators \cite{Kagan}. Models
with $Z$-mediated FCNC's \cite{ZFCNC} will affect $b\to s$ hadronic
penguins only marginally \cite{newphysics}. It should also be noted that if
the new physics in the $b\to s$ FCNC has the same phase as in the SM (i.e.\
$\simeq 0$), then the SM predictions will be unchanged. That is, there will
be no discrepancy in the value of $\beta$ extracted from
$\bdbarp\to\Psi\ks$ and from hadronic $b\to s$ penguins.

We now turn to an examination of the prospects for measuring $\beta$ via
$b\to s$ hadronic penguins. One needs a final state which is a CP
eigenstate. The decay mode first considered in this regard was
$\bdbarp\to\phi\ks$ \cite{LonPeccei}. However, there is another possibility
which has generated considerable excitement: $\bdbarp\to\eta'\ks$. The
reason that this mode is interesting is that the branching ratio for the
corresponding charged $B$ decay has recently been found to be anomalously
large: $B(B^+ \to \eta' K^+) = (7.8^{+2.7}_{-2.2} \pm 1.0) \times 10^{-5}$
\cite{CLEO}. We therefore expect that $B(\bd \to \eta' \ks) \simeq 4 \times
10^{-5}$. This large branching ratio greatly improves the usefulness of the
decay mode $\bdbarp\to\eta'\ks$ for getting at $\beta$.

In fact, there are a large number of final states that can be used:
$\phi\ks$, $\eta' \ks$, $\pi^0 \ks$, $\rho^0 \ks$, $\omega \ks$, $\eta
\ks$, etc. In principle, one can simply add the measured CP asymmetries in
all of these modes, including a minus sign if the CP of the final state is
negative, to obtain a larger signal.

There are two comments to be made here. First, although $\phi\ks$ is pure
$b\to s$ penguin, the other final states get some contributions from the
quark-level ${\bar b} \to {\bar u} u {\bar s}$ tree diagram. Since the tree
diagram has a different weak phase than that of the penguin, this can spoil
the cleanliness of the method for extracting $\beta$. In order to see how
important the effect is, one has to estimate the ratio of the tree ($T$)
and penguin ($P$) amplitudes.

Let us first consider $\bdbarp\to\eta'\ks$. We write
\beq
{T^{\eta' \ks} \over P^{\eta' \ks}} = 
{T^{\eta' \ks} \over T^{\pi^+\pi^-}} \,
{T^{\pi^+\pi^-} \over P^{\eta' \ks}} ~.
\eeq
The first piece on the right-hand side (RHS) can be estimated by a simple
comparison of the Feynman diagrams. The tree contribution to the decay $\bd
\to \pi^+ \pi^-$ is color-allowed and is controlled by the CKM matrix
elements $V_{ub}^* V_{ud}$. Compared to this, the analogous contribution to
$\bd \to \eta' \ks$ is suppressed by three factors: (i) it is colour
suppressed, (ii) it has the CKM matrix elements $V_{ub}^* V_{us}$, which is
a factor $\lambda$ smaller, and (iii) one has to include the normalizations
$u{\bar u} \sim \eta' / \sqrt{3}$ and $d {\bar s} \sim \ks / \sqrt{2}$.
There may be additional $SU(3)$-breaking effects involving form factors or
decay constants, but these are unknown and are probably of $O(1)$. The
second piece on the RHS can be obtained from the measured branching 
ratios, assuming that the decays $\bd \to \pi^+ \pi^-$ and $\bd \to \eta'
\ks$ are dominated by the tree and penguin contributions, respectively. The
upper limit on $\bd \to \pi^+ \pi^-$ is $1.5 \times 10^{5}$ \cite{CLEO}. We
thus obtain
\beq
{T^{\eta' \ks} \over P^{\eta' \ks}} < {\lambda \over \sqrt{6}} \, {a_2
\over a_1} \, \sqrt{1.5 \over 4} = 0.018 ~,
\eeq
where we have conservatively taken the color-suppression factor to be
$a_2/a_1 = 1/3$. (For $B$ decays into heavier final states, the suppression
factor has been found to be even smaller: $a_2/a_1 = 0.2$ \cite{Alam}.)
Thus the tree contribution to $\bd \to \eta' \ks$ is negligible, so that
the CP asymmetry in this mode does indeed measure $\beta$ to a very good
approximation.

The ratio of $T/P$ for the other decay modes can be calculated similarly.
Isospin implies that the amplitude for $\bd \to \pi^0 \ks$ is half that of
$\bd \to \pi^- K^+$. From the measured branching ratio of $B(B_d^0 \to \pi^-
K^+) = (1.5^{+0.5+0.2)}_{-0.4-0.2)}) \times 10^{-5}$ \cite{CLEO} one
therefore obtains $B(\bd \to \pi^0 \ks) \simeq 4 \times 10^{-6}$. The
remaining final states are not really related to $\pi^- K^+$, but one can
still estimate their branching ratios by comparing their penguin
contributions to that of $\pi^- K^+$. In this case, one finds that $B(\bd
\to \rho^0 \ks) \simeq B(\bd \to \omega \ks) \simeq 4 \times 10^{-6}$ and
$B(\bd \to \eta \ks) \simeq 1 \times 10^{-6}$. In all cases, we then obtain
\beq
{T \over P} \lsim 0.04 ~.
\eeq
Thus, these modes are also pure $b\to s$ penguin to a good approximation.

There is one detail which is worth mentioning. Due to $SU(3)$ breaking, the
physical $\eta$ and $\eta'$ are in fact linear combinations of the $\eta$
and $\eta'$ states used above. However, since in all cases the tree
contribution is very small, the inclusion of $\eta$-$\eta'$ mixing does not
affect the analysis.

To sum up this point, CP asymmetries in $b\to s$ penguins do indeed measure
the CP angle $\beta$. The tree contributions to these decays are quite
small, at most a few percent. It is therefore possible to add up the
measured CP asymmetries in all these modes to obtain a larger signal. If the
value of $\beta$ extracted in this way differs by more than about 10\% from
that found in $\Psi \ks$, then it is a clear signal of new physics, with
new phases, in the $b\to s$ FCNC. If the difference is less than about
10\%, it could in principle be due to the tree contamination. However, this
can be checked by using only the final states $\phi\ks$ and $\eta' \ks$ (to
a very good approximation).

The second comment concerns the standard way of extracting $\beta$ via the
CP asymmetry in $\bdbarp \to \Psi \ks$. Given that one can also obtain
$\beta$ through $b\to s$ penguins, how do these two methods compare with
one another?
\begin{itemize}

\item $\bdbarp\to\Psi\ks$: The branching ratio is $4 \times 10^{-4}$.
Assuming that the $\ks$ is detected in the $\pi^+ \pi^-$ mode (67\%
branching ratio), and that the $\Psi$ is detected via its decay to
$\mu^+\mu^-$ or $e^+e^-$ (12\% b.r.), the product branching ratio for
$\bdbarp\to\Psi\ks$ is $\sim 32 \times 10^{-6}$. 

\item $\bdbarp\to\eta'\ks$: The branching ratio for this process has been
found to be about $4 \times 10^{-5}$. The $\eta'$ has two important modes
through which it may be detected: $\eta' \rightarrow \eta\pi\pi$  (followed
by $\eta \rightarrow \gamma\gamma$) with a product b.r. of $17\%$, and
$\eta' \rightarrow \rho^0 \gamma$ (followed by $\rho^0 \rightarrow
\pi^+\pi^-$) with a product b.r.\ of about $30\%$. Thus the total
efficiency can approach $47\%$. Although CLEO's detection efficiency for
the $\eta'$ is at present only about $5\%$ (where so far only the $\eta \pi
\pi$ mode has been used), it is clearly important to improve upon this at
the $B$-factory detectors. For the purpose of our discussion we will assume
a combined efficiency of $20\%$, yielding an effective b.r.\ of $5 \times
10^{-6}$, which includes the b.r.\ for $\ks \to \pi^+\pi^-$.

\item $\bdbarp\to\phi\ks$: The branching ratio for this decay has not yet
been measured, but we can estimate its value from measured quantities. The
decay $B_d^0 \to \pi^- K^+$, which has a branching ratio of $1.5 \times
10^{-5}$, should be in the same ballpark as $\bdbarp\to\phi\ks$. Taking
into account that $d {\bar s} \sim \ks / \sqrt{2}$, we therefore estimate
its branching ratio to be in the range of $(7-15) \times 10^{-6}$. (This is
slightly larger than theoretical estimates which also include electroweak
penguin contributions \cite{Fleischer}.) Assuming that the $\phi$ is
detected through its $K^+ K^-$ decay mode (49\% b.r.), the product
branching ratio for this decay is $\sim (2-5) \times 10^{-6}$.

\end{itemize}
The remaining modes --- $\pi^0 \ks$, $\rho^0 \ks$, $\omega \ks$, $\eta \ks$
--- can be analyzed similarly, using the decays $\pi^0 \to \gamma\gamma$
(99\% b.r.), $\rho^0 \to \pi^+\pi^-$ (100\% b.r.), $\omega \to
\pi^+\pi^-\pi^0$ and $\pi^+ \pi^-$ (91\% b.r.), and $\eta \to \gamma\gamma$
(39\% b.r.). In this case one needs estimates of the branching ratios and
of the detection efficiencies for the neutral mesons, which at this point
are unknown. A reasonable educated guess is that all of these product
branching ratios are in the range $\sim (2-5) \times 10^{-6}$. Thus, when
all the $b\to s$ penguin decay modes are added, they could have a combined
yield of up to $3\times 10^{-5}$. This approach is therefore quite
promising: it may end up requiring just about the same number of $B$'s as
compared to $\Psi \ks$, or perhaps a factor of two or three more.

To conclude, our observations are quite simple. $B$ factories are going to
measure CP-violating asymmetries using $\bd$ and $B^\pm$ decays. Although
there are a variety of ways of detecting the presence of physics beyond the
SM, such measurements are sensitive {\it only} to new physics in
$\bd$-$\bdbar$ mixing, i.e.\ in the $b\to d$ FCNC. However, it is
conceivable that the new physics enters only in the $b\to s$ FCNC, leaving
the $b\to d$ FCNC unaffected. Since $B$ factories are not going to be able
to measure CP asymmetries in $\bs$ decays, it appears that this possibility
will remain untested. 

In fact, there {\it is} a way of probing new physics in the $b\to s$ FCNC.
In the SM, CP asymmetries in $\bdbarp\to\Psi\ks$ and $b\to s$ hadronic
penguins both measure the CP angle $\beta$. If there is a difference
between the measurements of $\beta$ obtained in these two ways, this
directly indicates the presence of new physics, with new phases, in the
$b\to s$ FCNC. In general, this same new physics will affect $\bs$-$\bsbar$
mixing, and will shown up in CP asymmetries involving $\bs$ decays. 

Recent CLEO results are encouraging in this regard. In particular, there is
evidence that the branching ratio for the decay $B_d^0 \to \eta'\ks$ is
anomalously large, about $4 \times 10^{-5}$. The CP asymmetry measured in
this decay probes the CP angle $\beta$ to a very good approximation. One
can also measure $\beta$ via the CP asymmetries in $\bd$ decays to
$\phi\ks$, $\pi^0 \ks$, $\rho^0 \ks$, $\omega \ks$, $\eta \ks$, etc. The
branching ratios for these decays have not yet been measured, but are
expected to be $\sim 5 \times 10^{-6}$. In principle it should be possible
to add up the measured CP asymmetries in all these modes to obtain a larger
signal. The number of $B$'s required to measure $\beta$ using this method
may be about the same, or perhaps only a factor of 2-3 more, as that needed
in the conventional mode, $\bdbarp\to\Psi\ks$. In any case, it is clear
that the branching ratios for $b\to s$ hadronic penguins are sufficiently
large that the CP asymmetries in these decays can probably be measured as
easily as those used for the extraction of the other CP angles $\alpha$ and
$\gamma$. With this in mind, it will be important to maximize the detection
efficiency for the $\eta'$.

This method is perhaps the only way of detecting the presence of new phases
in the $b\to s$ FCNC without measuring CP asymmetries in $\bs$ decays, and
therefore should be a high priority at future $B$ factories.

\bigskip
\centerline{\bf Acknowledgements}
\bigskip
We thank Jim Alexander, Bruce Behrens, Tom Browder, Peter Kim and Sheldon
Stone for discussions. DL is grateful for the pleasant hospitality of the
Brookhaven National Laboratory, where much of this work was done. This
research was financially supported by NSERC of Canada and FCAR du Qu\'ebec
and in part by U.S. DOE contract DE-AC-76CH00016.

\end{document}